\documentclass[a4paper,11pt]{article}
\usepackage{pos}
\usepackage{subfig}

\title{Modified lepton couplings and the Cabibbo-angle anomaly}

\author*[a,b]{Claudio Andrea Manzari}

\author[b]{Antonio M. Coutinho}
\author[a,b,c]{Andreas Crivellin}

\affiliation[a]{Universit\"{a}t Z\"{u}rich,\\
 Winterthurerstrasse 190, CH--8057 Z\"{u}rich, Switzerland}

\affiliation[b]{Paul Scherrer Institut,\\
 CH--5232 Villigen PSI, Switzerland}

\affiliation[c]{CERN Theory Division,\\
 CH--1211 Geneva 23, Switzerland}

\emailAdd{claudioandrea.manzari@physik.uzh.ch}
\emailAdd{antonio.coutinho@psi.ch}
\emailAdd{andreas.crivellin@cern.ch}

\abstract{Significant discrepancies between the different determinations of the Cabibbo angle have been observed. Here, we point out that this “Cabibbo-angle anomaly” can be explained by lepton flavour universality (LFU) violating New Physics (NP) in the neutrino sector. However, modified neutrino couplings to Standard Model gauge bosons also affect many other observables sensitive to LFU violation, which have to be taken into account in order to assess the viability of this explanation. Therefore, we perform a model-independent Bayesian global analysis and find that non-zero modifications of electron and muon neutrino couplings are preferred at more than $99.99\%\,{\rm C.L.}$  (corresponding to more than $4\,\sigma$). Our results show that constructive effects in the muon sector are necessary, meaning simple models with right-handed neutrinos are discarded and more sophisticated NP models required.}

\FullConference{The Eighth Annual Conference on Large Hadron Collider Physics-LHCP2020 \\
  25-30 May, 2020\\
  online}

\begin{document}
\maketitle

\section{Introduction}

\begin{figure}[h!]
	\centering
	\includegraphics[width=0.54\textwidth]{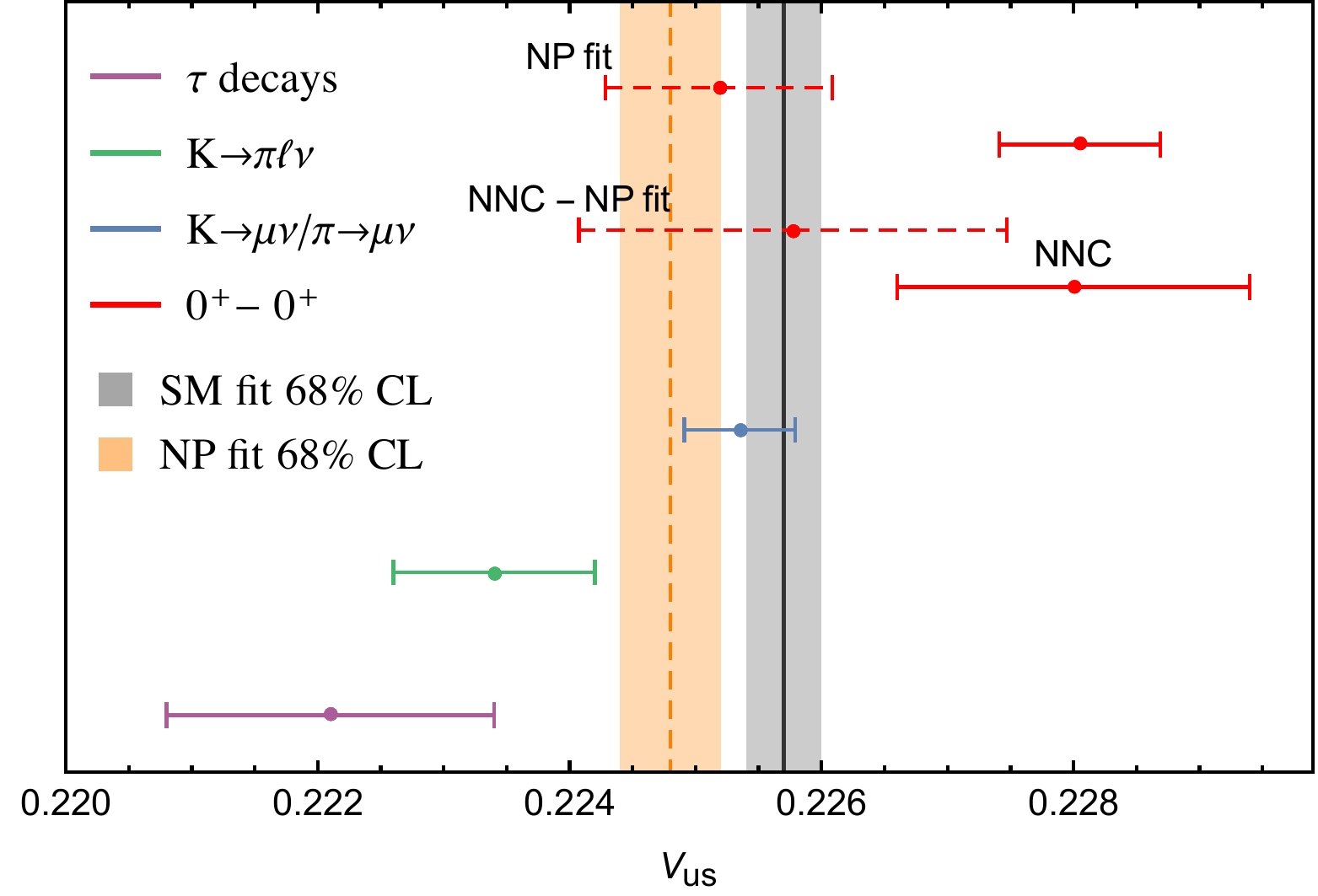}
	\caption{Measurements of $V_{us}$ from $\tau$ decays, $K\to \pi\ell\nu$, ${K\to\mu\nu}/{\pi\to\mu\nu}$, and $0^+ - 0^+$ transition using CKM unitarity to convert $V_{ud}$ to $V_{us}$. The grey band shows the 68\% CL posterior within the SM while the orange band corresponds to the NP fit with non-zero values of $\varepsilon_{ii}$. Accidentally, the posterior of $V_{us}$ is the same, independently of the theory input used for beta decays (to the numerical accuracy at which we are working). The effect of $\varepsilon_{ii}$ on $V_{us}$ within the NP fit is quite small except for the determination from super-allowed beta decay.}
	\label{fig:Anomaly}
\end{figure}

The global electroweak (EW)~\cite{deBlas:2016ojx,Haller:2018nnx,Aaltonen:2018dxj} and the CKM~\cite{Ciuchini:2000de,Hocker:2001xe} fits contributed to establish the Standard Model (SM) of particle physics with an extraordinary precision in the last decades. However, at present, there are tensions between the different determinations of the Cabibbo Angle from the CKM elements $V_{us}$ and $V_{ud}$, known as the ``Cabibbo Angle Anomaly'' (CAA). In detail, the different determinations of $V_{us}$ are from:
\vspace{-1mm}
\begin{itemize}
	\item Measurements of $K\to\pi\ell\nu$ together with the form factor $f_+(0)$ evaluated at zero momentum transfer result in $V_{us}=0.2232(11)$~\cite{Aoki:2019cca}. 
	\vspace{-1.5mm}
	\item $K\to\ell\nu/\pi\to\ell\nu$ where $V_{us}/V_{ud}$ is determined once the ratio of the decay constants $f_{K^{\pm}}/f_{\pi^{\pm}}$ is known. Using CKM unitarity this results in $V_{us}=0.22534(44)$~\cite{Aoki:2019cca}. 
	\vspace{-1.5mm}
	\item $V_{ud}$ measured in super-allowed nuclear $\beta$ decay~\cite{Hardy:2017G0}, with $V_{us}$ again determined via CKM unitarity. Here, the description relies heavily on the evaluation of radiative corrections and we consider the latest estimates in Ref.~\cite{Seng:2020wjq} (NNC indicate the scenario with the nuclear corrections discussed for the first time in~\cite{Seng:2018qru,Gorchtein:2018fxl}): 
	\begin{align}
	|V_{us}| = 0.22805(64) \qquad	\qquad |V_{us}|_{\rm NNC} = 0.2280(14)
	\end{align}
	\item $\tau\to K\nu$, $\tau\to K\nu/\tau\to\pi\nu$ and inclusive tau decays which provide additional measurements of $V_{us}$ and $V_{us}/V_{ud}$. Here the HFLAV average is $V_{us}=0.2221(13)$~\cite{Amhis:2019ckw}.
\end{itemize}
\vspace{-1mm}
This situation is graphically depicted in Fig.~\ref{fig:Anomaly}. One can clearly see that these measurements are not consistent with each other, and Ref.~\cite{Seng:2020wjq} quantifies this inconsistency to be at the level of $3-5~\sigma$, or $1.7-3~\sigma$ if NNC are included.

This anomaly was studied in the scenario of first row CKM unitarity violation~\cite{Belfatto:2019swo,Cheung:2020vqm}, however it is challenging to explain it in this context due other flavour bounds, like Kaon or $B_d-\bar B_d$ mixing~\cite{Bobeth:2016llm}. In this work, driven by the observation that all the processes discussed above involve electron and muon neutrinos, we study this anomaly as an evidence of Lepton Flavour Universality Violation (LFUV), opening the possibility for connections with the hints for NP in semi-leptonic $B$ decays and the anomalous magnetic moments of the muon and electron~\cite{Crivellin:2020lzu, Lees:2012xj,Aaij:2017deq,Abdesselam:2019dgh,Aaij:2017vbb,Aaij:2019wad,Bennett:2006fi,Aoyama:2020ynm,Davoudiasl:2018fbb,Crivellin:2018qmi,Kirk:2020wdk}.

\section{Setup}

In this work we consider a scenario which can lead to modifications in the processes discussed so far: modified neutrino couplings to gauge bosons. We study this scenario in an EFT approach assuming that the NP scale is above the EW scale as suggested by LHC\cite{Butler:2017afk,Masetti:2018btj} and LEP~\cite{Antonelli:2001rz} searches. At the dimension-6 level, there is just one operator which modifies only the couplings of neutrinos to gauge bosons~\cite{Buchmuller:1985jz,Grzadkowski:2010es}, ${\bar L_i}{\gamma^\mu }\tau _{}^I{L_j}{H^\dag }i{\mathord{\buildrel{\lower3pt\hbox{$\scriptscriptstyle\leftrightarrow$}} 
		\over D}_\mu^I}H$, with 	${\tau ^I} = \left( {1, - {\sigma _1}, - {\sigma _2}, - {\sigma _3}} \right)$ 
where $\sigma_i$ are the Pauli matrices (it is the difference of the operators $Q_{\phi\ell}^{(1)}$ and $Q_{\phi\ell}^{(3)}$ in the basis of Ref.~\cite{Grzadkowski:2010es}). In the following, we conveniently parametrize the effect of the Wilson coefficient of this operator in such a way that the Feynman rules for the $W$ and $Z$ couplings with neutrinos become, respectively
\begin{align}
 \frac{{ - i{g_2}}}{{\sqrt 2 }}{{\bar \ell }_i}{\gamma ^\mu }{P_L}{\nu _j}{W_\mu }\left( {{\delta _{ij}} + \frac{1}{2}{\varepsilon _{ij}}} \right)\,,\qquad \frac{{ - i{g_2}}}{2c_W}{{\bar \nu }_i}{\gamma ^\mu }{P_L}{\nu _j}{Z_\mu }\left( {{\delta _{ij}} + {\varepsilon _{ij}}} \right)\,.
\label{couplings}
\end{align}

As already outlined in the previous sections, we want to study whether modified neutrino couplings could provide a valid explanation for the CAA anomaly. However, they directly and indirectly modify other two important classes of physical observables: the EW observables measured with high precision at LEP, Tevatron and LHC, and the low energy observables testing LFU. Correlated effects may arise, and a global fit to all data is necessary to assess consistently the full impact of our scenario. For a complete and detailed discussion on our global fit we refer the interested reader to~\cite{Coutinho:2019aiy}, while here we give a concise overview. Note that the non-diagonal elements of $\varepsilon_{ij}$ in Eq.~\ref{couplings} will be safely neglected in what follows because of the very stringent constraints from radiative lepton decays~\cite{TheMEG:2016wtm,Aubert:2009ag}  and since they enter only quadratically in flavour conserving processes.

We include the usual set of EW observables measured with high precision at LEP~\cite{ALEPH:2005ab,Schael:2013ita}. The EW sector of the SM can be completely parametrized by 3 Lagrangian parameters and we choose the set with the smallest experimental error:  $\alpha$, $M_Z$ and $G_F$. Among these parameters, only the determination of  the latter~\cite{Webber:2010zf} (which is determined with a very high precision from the muon lifetime) is affected by the modification of the neutrino couplings in Eq.~\ref{couplings}. Denoting $G_F^{\mathcal{L}}$ the parameter appearing in the lagrangian, we find for the measured quantity
\begin{align}
\begin{split}
G_F^{}&=G_F^{\mathcal{L}} \left(1+\frac{1}{2}\varepsilon_{ee}+\frac{1}{2}\varepsilon_{\mu\mu}\right)\,.
\end{split}
\label{GFmod}
\end{align}
As a consequence, in addition to the direct modifications of W and Z decay observables, most of the observables in the EW sector are indirectly modified from the shift in $G_F$.

In case the modified neutrino couplings violate lepton flavour universality (LFU), the low energy observables testing LFU provide stringent constraints. Here, we have ratios of W decays~\cite{Alcaraz:2006mx} ($W\to \ell_i\nu_i /W\to\ell_j\nu_j$) where we included the latest ATLAS results~\cite{Aad:2020ayz}, as well as of Kaon~\cite{Ambrosino:2009aa,Lazzeroni:2012cx}, pion~\cite{Aguilar-Arevalo:2015cdf} and tau decays (see Ref.~\cite{Pich:2013lsa} for a recent overview). 

For the determinations of $V_{us}$ we consider semi-leptonic Kaon decays, $V_{us}^{K_{\mu3}}$, the ratio between Kaon and pion leptonic decays, $V_{us}^{K/\pi}$ , inclusive $\tau$ decays, $V_{us}^{\tau}$, and superallowed beta decays, $V_{us}^{\beta}$. $V_{us}^{K/\pi} $ and $V_{us}^{\tau}$ are not affected by modified neutrino couplings while one find the following relations between the Lagrangian value, $V_{us}^{\mathcal{L}}$, and the measured $V_{us}^{K_{\mu3}}$ and $V_{us}^{\beta}$ (including $G_F$ indirect effects)
\begin{align}
\begin{split}
|V_{us}^{K_{\mu3}}| \simeq |V_{us}^{\mathcal{L}}|\bigg(1-\frac{1}{2}\varepsilon_{ee}\bigg)\,, \qquad
|V_{us}^{\beta}| \simeq \sqrt{1-|V_{ud}^{\mathcal{L}}|^2(1-\frac{1}{2}\varepsilon_{\mu\mu})^2}\,.
\end{split}
\end{align}

\section{Results and Conclusions}

\begin{figure}[h!]
	\centering
\includegraphics[width=0.56\textwidth]{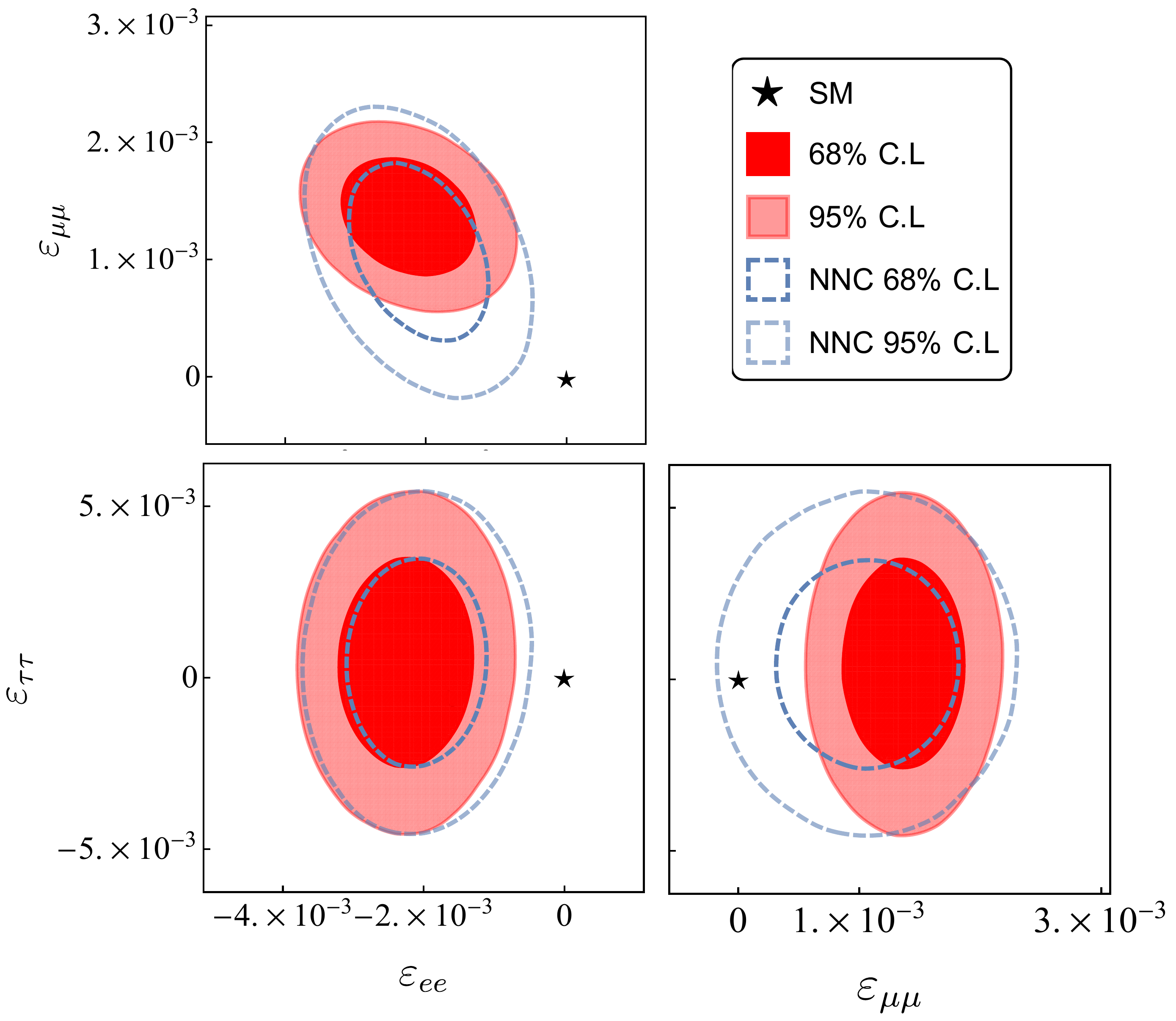}
	\caption{Result of the Global Fit using the estimates for $V_{us}^{\beta}$ with and without NNC. The 2-D fit ($68\%$ and $95\%$ C.L) for $\varepsilon_{ii}-\varepsilon_{jj}$ are shown.}
	\label{fig:ResultFit}
\end{figure}

Our analysis is performed in a Bayesian framework using the \texttt{HEPfit} package~\cite{deBlas:2019okz}. Concerning the parameters of the fit, for $G_F,\,\alpha,\,\alpha_s,\,M_Z,\,m_t$ and $m_H$ we assume Gaussian priors, corresponding to the current direct measurements or evaluations, whereas $V_{us}^\mathcal{L}$ and $\varepsilon_{ii}$ have flat priors. In Fig.~\ref{fig:ResultFit} we show the result of the fit for the two dimensional $\varepsilon_{ii}-\varepsilon_{jj}$ planes. The $68\%$ and $95\%$ C.L. contours for the two scenarios for $V_{us}^{\beta}$ discussed above are shown. It is clear from the $\varepsilon_{ee}-\varepsilon_{\mu\mu}$ plane, where the largest deviation from SM can be found, that these regions do not overlap with the SM point $\varepsilon_{ii}=0$, and that $\varepsilon_{ee}$ and $\varepsilon_{\mu\mu}$ possess an anti-correlation and an opposite sign. For a more direct model comparison between the NP hypothesis and the SM we look at values of the Information Criterion (IC)~\cite{Ando:2007,Ando:2011}, a quantity characterized by an estimate of the predictive accuracy of the fitted model~\cite{Gelman:2013} and a penalty factor for its number of free parameters. Here we obtain: for the SM, $\text{IC}_{\rm SM}\simeq93$, compared to $\text{IC}_{\rm NP}\simeq\text{IC}_{\rm NP-NNC}\simeq76$ for the two NP scenarios. In the vein of Refs.~\cite{Jeffreys:1998,Kass:1995}, this constitutes "very strong" evidence against the SM. In Fig.~\ref{fig:Anomaly} we also show the posterior for $V_{us}$ and the value extracted from super-allowed beta decay using the parameters fitted. This result shows that current data clearly favours the NP hypothesis, exclude conventional models with right-handed neutrinos leading to necessary destructive interference~\cite{Mohapatra:1986bd,BERNABEU1987303,BRANCO1989492,BUCHMULLER1990458,Pilaftsis:1991ug,Malinsky:2005bi,Dev:2012sg,Antusch:2014woa,Coy:2018bxr} and promote the search for NP models as the ones presented in Refs.~\cite{Capdevila:2020rrl,Crivellin:2020ebi}.

\newpage

\bibliographystyle{JHEP} 
\bibliography{bibliography}

\end{document}